\documentstyle[aaspp4]{article}
\begin{document}

\title{$^{13}$CO($J$ = 1 -- 0) DEPRESSION IN LUMINOUS STARBURST MERGERS}

\author{Yoshiaki Taniguchi and Youichi Ohyama}

\affil{Astronomical Institute, Tohoku University, Aoba, Sendai 980-8578, Japan}


\begin{abstract}
It is known that the class of luminous starburst galaxies tends to have higher
$R =^{12}$CO($J$=1--0)$/^{13}$CO($J$=1--0) integrated line intensity ratios
($R>20$) than normal spiral galaxies ($R\sim 10$).
Since most previous studies investigated only $R$, it remains uncertain
whether the luminous starburst galaxies are overabundant in $^{12}$CO or
underabundant in $^{13}$CO.
Here we propose a new observational test to examine this problem.
Our new test is to compare far-infrared luminosities
[$L$(FIR)] with those of $^{12}$CO and $^{13}$CO
[$L$($^{12}$CO) and $L$($^{13}$CO), respectively].
It is shown that there is a very tight correlation between $L$($^{12}$CO) and $L$(FIR),
as found in many previous studies. However, 
we find that the $^{13}$CO luminosities of the high-$R$
galaxies are lower by a factor of three on the average than
those expected from the correlation for the remaining galaxies with
ordinary $R$ values.
Therefore, we conclude that the observed high $R$ values for the luminous 
starburst galaxies are attributed to their lower  $^{13}$CO line intensities.
\end{abstract}




\section{INTRODUCTION}

It is known that the class of luminous starburst galaxies
with $L$(FIR)$\gtrsim 10^{11}L_{\odot}$ tends to have higher
$R=^{12}$CO($J$=1--0)$/^{13}$CO($J$=1--0) integrated line intensity ratios
than normal spiral galaxies (Aalto et al. 1991, 1995, 1997;
Casoli et al. 1991;
Casoli, Dupraz, \& Combes 1992a, 1992b; Hurt \& Turner 1991;
Turner \& Hurt 1992; Garay, Mardones, \& Mirabel 1993;
Henkel \& Mauersberger 1993; Henkel et al. 1998).
It was often considered that the higher $R$ values observed in these galaxies
are attributed to the inflow of disk gas 
with high $^{12}$C$/^{13}$C abundance ratios,
possibly combined with a $^{12}$C enhancement caused by nucleosynthesis
in massive stars (e.g., Henkel et al. 1998).
However, since most previous studies investigated only the ratios,
it is uncertain whether or not these luminous starburst galaxies 
are overluminous in $^{12}$CO or underluminous in $^{13}$CO (Casoli et al. 1992a).
In this Letter, comparing the luminosities of $^{12}$CO($J$=1--0), 
$^{13}$CO($J$=1--0) (hereafter $^{12}$CO and $^{13}$CO, respectively), and
FIR emission for a sample of starburst and normal galaxies for
which all the data are available in literature,
we show that the observed higher $R$ values are mainly attributed to
a lower intensity in $^{13}$CO with respect to $^{12}$CO.
We discuss possible physical mechanisms which can explain the 
$^{13}$CO($J$=1--0) depression in luminous starburst galaxies.


\section{DATA AND RESULTS}

We have compiled $^{12}$CO and $^{13}$CO intensities
from the literature (Wiklind \& Henkel 1990; Aalto et al. 1991, 1995;
Becker \& Freudling 1991; Sage \& Isbell 1991; Casoli et al. 1992b;
Garay et al. 1993; Henkel, Whiteoak, \& Mauersberger 1994).
Our sample consists of 51 galaxies including ultraluminous infrared galaxies
such as Arp 220.
These data are used to estimate the CO luminosities;
$L$(CO) is defined as $L$(CO)$=A\times I$(CO) K km s$^{-1}$ pc$^2$
where $A$ is the observed area in units of pc$^2$ and
$I$(CO)$=\int T_{\rm A}^* \eta^{-1} dv$ K km s$^{-1}$
where $T_{\rm A}^*$ is the observed antenna temperature
corrected for atmospheric extinction and $\eta$ is the main beam efficiency.

The FIR data are compiled from the IRAS Faint Source Catalog
(Moshir et al. 1992).
The FIR luminosities are estimated using
$L$(FIR)$=4\pi D^2
1.26\times 10^{-11} [2.58\times S(60) + S(100)]$ (ergs s$^{-1}$)
where $S$(60) and $S$(100) are the IRAS 60 $\mu$m and 100 $\mu$m fluxes
in units of Jy and $D$ is the distance of galaxies in units of cm
(Helou, Soifer, \& Rowan-Robinson 1985).
Distances of nearby galaxies are taken from the Nearby Galaxies Catalog
(Tully 1988); distances of other galaxies are estimated using a Hubble constant
$H_0$ = 75 km s$^{-1}$ Mpc$^{-1}$ with $V_{\rm GSR}$ (radial velocity 
corrected to the Galactic Center: de Vaucouleurs et al. 1991).
The compiled data are given in Table 1.
Though our sample is not statistically complete,
it is the largest sample compiled so far.

As shown in Table 1, there are several galaxies which have $R \geq 20$.
Hereafter we refer to these sources as high-$R$ galaxies (NGC 1614, NGC 3256,
NGC 4194, NGC 6240, Arp 220, and Arp 299).
ESO 541--IG 23 [= Arp 236, $R\geq 17$ (lower limit)] and possibly
IRAS 18293$-$3413 ($R\simeq 26$: Garay et al. 1993;
$R\simeq 17$: Aalto et al. 1995) are also high-$R$ galaxies.
In total, there are eight high-$R$ galaxies in our sample.
Except for the uncertain case of IRAS 18293$-$3413, all these high-$R$ galaxies
are mergers.
For the remaining 43 galaxies, we obtain an average ratio,
$<R> = 11.3\pm3.3$.
The values are consistent with  previous estimates of $11\pm 3$
(Aalto et al. 1991) and $9.3\pm3.6$ (Sage \& Isbell 1991)
for normal galaxies.

First, we examine whether or not there is a beam size effect on $R$
because the CO line observations did not cover the whole area of the individual
galaxies.
Ratios of beam size ($D_{\rm beam}$ = half power beam width) to
optical diameter at 25 mag arcsec$^{-2}$ ($D_{25}$:
de Vaucouleurs et al. 1991) are given in the third column of Table 1.
In Figure 1, $D_{\rm beam}/D_{25}$ is plotted as a function of $R$.
We find no clear correlation, indicating that the difference in $R$ cannot be
attributed to the beam-size effect.

In Figure 2, we compare $L$($^{12}$CO) with $L$($^{13}$CO).
A tight correlation is found for
galaxies with $L$($^{12}$CO)$<10^{8.75}$ K km s$^{-1}$ pc$^2$.
This almost linear correlation can be expressed as
log $L$($^{13}$CO)$=$($0.96\pm0.03$)$\times$ log
$L$($^{12}$CO)$+$($-0.68\pm0.27$).
However, more luminous galaxies do not follow the same correlation,
i.e., some of the luminous galaxies have lower $^{13}$CO luminosities
with respect to $^{12}$CO.

In Figure 3, we compare $L$(FIR) with both $L$($^{12}$CO) and $L$($^{13}$CO).
It is shown that $L$($^{12}$CO) is well correlated with $L$(FIR);
its correlation coefficient is 0.94. The average $L(^{12}$CO)/$L$(FIR)
for the high-$R$ galaxies, 0.013$\pm$0.007, is similar to that for 
the normal ones, 0.023$\pm$0.013.
This correlation has been noted in many previous studies
(e.g., Young \& Scoville 1991 and references therein).
In fact, Figure 3 shows that the correlation is established within
a factor of three dispersion except for one galaxy (NGC 55) 
which is a nearby Magellanic-type irregular galaxy (Becker \& Freudling 1991).
On the other hand, the correlation between
$L$(FIR) and $L$($^{13}$CO) appears to be significantly poorer
(the correlation coefficient is 0.86) than that between $L$(FIR) 
and $L$($^{12}$CO). The correlation is poorer because
the high-$R$ luminous starburst galaxies 
have lower $^{13}$CO luminosities than expected
for galaxies with ordinary $R$ values; note that the average value of
$L(^{13}$CO)/$L$(FIR) is 0.00051 $\pm$ 0.00034 for the high-$R$ galaxies
while that for the normal galaxies is 0.0023 $\pm$ 0.0017.
Thus, we conclude that the higher $R$ in the luminous
starburst mergers is mostly due not to enhanced $^{12}$CO
emission but to depressed $^{13}$CO line intensities.
An average $R$ value for the high-$R$ galaxies is $\simeq 30.7 \pm 11.5$.
Since lower-limit data are used in this estimate,
we obtain a $^{13}$CO depression factor $\gtrsim 3$,
given $<R> \simeq 11$ for the normal galaxies. 
Any physical mechanism must explain the intrinsic 
weakness of $^{13}$CO($J$ = 1 -- 0) emission by a factor of $\gtrsim 3$ 
in the luminous starburst mergers.

\section{DISCUSSION}

Here we investigate possible mechanisms that can explain the intrinsic
weakness of $^{13}$CO($J$ = 1 -- 0) emission by a factor of $\gtrsim 3$
in the luminous starburst mergers.
High $R$ values may be caused by (see also Sakamoto et al. 1997):
1) low gas densities, 
2) high gas kinetic temperatures,
3) large nuclear velocity gradients that keep CO column densities 
small at a given velocity,
4) small CO/H$_2$ abundance ratios, 
5) more efficient excitation of $^{13}$CO molecules toward upper levels
(e.g., $J \geq 2$) with respect to $^{12}$CO ones, or 
6) large $^{12}$CO/$^{13}$CO abundance ratios.
The first four mechanisms lead to lower opacity of the CO lines.
The first mechanism also introduces non-LTE excitation, resulting in 
high $R$ values.
Since it is known that the luminous starburst mergers have a huge amount
of dense gas ($\sim 10^{10} M_{\odot}$; Solomon et al. 1992; 
Scoville, Yun, \& Bryant 1997),
the first idea (lower gas densities) seems to be unlikely.
Starburst galaxies often have  both
high gas kinetic temperatures and particularly large velocity gradients
(Turner, Martin, \& Ho 1990; Aalto et al. 1991, 1995; Devereux et al. 1994).
However, if these mechanisms were responsible for the higher $R$ values, 
we would find a $^{12}$CO excess with respect to $^{13}$CO.
Therefore the second and third idea are also ruled out.
The smaller CO/H$_2$ abundance ratio (i.e., lower metallicity) seems also
to  be unlikely
for the luminous starburst mergers because the enhanced star-formation activity
occurs in their nuclear regions where metal abundances are  generally high
(e.g., Storchi-Bergmann et al. 1996).
It is known that $^{12}$CO($J$=3--2)/$^{12}$CO($J$=1--0) ratios of starburst galaxies
are often found to be higher than in normal galaxies (e.g., Devereux et al. 1994
and references therein). If $^{13}$CO molecules in starburst mergers 
were excited more efficiently toward
upper levels with respect to $^{12}$CO, we would observe  
high-$R$ values in higher rotational transitions. 
However, there is no such evidence (Taniguchi, Ohyama, \& Sanders 1998).
Therefore, the remaining possibility is that the $^{12}$CO/$^{13}$CO
abundance ratios  are systematically higher than 
those in normal galaxies.
According to our result (Fig. 3), this should be achieved by 
a $^{13}$CO underabundance with respect to $^{12}$CO in the 
starburst mergers.

\vspace{1ex}

We would like to thank Dave Sanders, Naomasa Nakai, Toshihiro Handa, 
Sumio Ishizuki, Naomi Hirano, and Satoki Matsushita for useful comments.
We would also like to thank the referee, Christian Henkel, for his many
useful comments and suggestions which improved this paper significantly.
Y.O. was supported by the Grant-in-Aid for JSPS Fellows by
the Ministry of Education, Science, Sports and Culture.
This work was supported in part by the Ministry of Education, Science,
Sports and Culture in Japan under Grant Nos. 07055044, 10044052, and 10304013.


\begin{deluxetable}{lccccccc}
\tablewidth{38pc}
\tablecaption{Properties of a sample of galaxies}
\tablehead{
\colhead{Name} &
\colhead{$d$} &
\colhead{$D_{\rm beam}$} &
\colhead{$L$($^{12}$CO)} &
\colhead{$L$($^{13}$CO)} &
\colhead{$R$} &
\colhead{$L$(FIR)} &
\colhead{Ref.\tablenotemark{a}} \\
\colhead{} &
\colhead{(Mpc)} &
\colhead{/$D_{25}$} &
\colhead{(K km s$^{-1}$ pc$^{2}$)} &
\colhead{(K km s$^{-1}$ pc$^{2}$)} &
\colhead{} &
\colhead{($L_{\sun}$)} &
\colhead{}
}
\startdata
N55  & 1 & 0.023 & 5.51 & 4.33 & 15.4 & 8.03 & 1 \nl
N134 & 19 & 0.084 & 8.29 & 7.14 & 14.0 & 10.16 & 2 \nl
N253 & 3 & 0.034 & 8.26 & 7.04 & 16.6 & 10.03 & 3 \nl
N404 & 2 & 0.106 & 5.75 & 4.79 & 9.1  & 7.33 & 4 \nl
N520 & 28 & 0.123 & 8.76 & 7.71 & 11.0 & 10.58 & 5 \nl
N628 & 10 & 0.091 & 7.57 & 6.75 & 6.6  & 9.64\tablenotemark{b} & 3 \nl
N660 & 12 & 0.066 & 8.58 & 7.43 & 14.1 & 10.17 & 5 \nl
N828 & 73\tablenotemark{c} & 0.191 & 9.73 & 8.70 & 10.7 & 11.03 & 5 \nl
\nodata & \nodata & 0.127 & 9.61 & 8.44 & 14.9 & \nodata & 6 \nl
N891 & 10 & 0.070 & 8.40 & 7.48 & 8.4  & 9.90 & 3 \nl
N986 & 23 & 0.184 & 8.75 & 7.74 & 10.2 & 10.35 & 5 \nl
N1058 & 9 & 0.315 & 7.13 & 6.07 & 11.5 & 8.70 & 3 \nl
N1614 & 62\tablenotemark{c} & 0.544 & 9.11 & $<$7.67 & $>$27.4 & 11.25 & 5 \nl
N1808 & 11 & 0.111 & 8.46 & 7.25 & 16.5 & 10.22 & 5 \nl
N2146 & 17 & 0.091 & 8.82 & 7.74 & 12.0 & 10.78 & 5 \nl
N2276 & 37 & 0.195 & 8.66 & 7.75 & 8.0  & 10.50 & 2 \nl
N2369 & 44\tablenotemark{c} & 0.202 & 9.39 & 8.22 & 15.0 & 10.82 & 2 \nl
\nodata & \nodata & 0.211 & 9.33 & 8.28 & 11.3 & \nodata & 7 \nl
N2903 & 6 & 0.075 & 7.79 & 6.81 & 9.5  & 9.48 & 3 \nl
N3034 & 5 & 0.085 & 8.67 & 7.47 & 15.9 & 10.62 & 3 \nl
N3079 & 20 & 0.069 & 8.94 & 7.91 & 10.8 & 10.52 & 5 \nl
N3256 & 37 & 0.189 & 9.57 & 8.01 & 35.6 & 11.28 & 5 \nl
\nodata & \nodata & 0.193 & 9.54 & 8.07 & 29.2 & \nodata & 1 \nl
\nodata & \nodata & 0.096 & 8.95 & 7.40 & 35.5 & \nodata & 6 \nl
\nodata & \nodata & 0.197 & 9.53 & 8.12 & 25.9 & \nodata & 7 \nl
N3627 & 7 & 0.104 & 7.94 & 6.77 & 14.8 & 9.57 & 3 \nl
N3628 & 8 & 0.064 & 8.34 & 7.32 & 10.5 & 9.73 & 3 \nl
N4038/9 & 25 & 0.086\tablenotemark{d} & 8.94 & 7.82 & 13.3 & 10.64 & 5 \nl
N4194 & 39 & 0.201 & 8.83 & 7.09 & 54.4 & 10.69 & 6 \nl
N4414 & 10 & 0.262 & 8.27 & 7.47 & 6.3  & 9.70 & 3 \nl
N4736 & 4 & 0.085 & 7.29 & 6.37 & 8.3  & 9.33\tablenotemark{b} & 3 \nl
N4826 & 4 & 0.055 & 7.22 & 6.51 & 5.1  & 9.03 & 5 \nl
\nodata & \nodata & 0.095 & 7.50 & 6.77 & 5.3  & \nodata & 3 \nl
N4945 & 5 & 0.036 & 8.61 & 7.45 & 14.4 & 10.22 & 8 \nl
\nodata & \nodata & 0.036 & 8.63 & 7.45 & 15.1 & \nodata & 5 \nl
N5033 & 19 & 0.051 & 8.27 & 7.32 & 8.9  & 10.04 & 5 \nl
N5055 & 7 & 0.044 & 7.65 & 6.86 & 6.2  & 9.57 & 5 \nl
N5218 & 39\tablenotemark{c} & 0.302 & 9.13 & 8.16 & 9.3  & 10.28 & 5 \nl
N5457 & 5 & 0.074 & 8.53\tablenotemark{e} & 7.69\tablenotemark{e} & 6.9\tablenotemark{e} & 
9.74\tablenotemark{b} & 3 \nl
N6215 & 21 & 0.335 & 8.21 & 7.13 & 12.1 & 10.33\tablenotemark{f} & 2 \nl
N6221 & 19 & 0.202 & 8.60 & 7.35 & 18.0 & 10.49\tablenotemark{f} & 2 \nl
N6240 & 98\tablenotemark{c} & 0.172 & 9.72 & 8.08 & 43.6 & 11.52 & 6 \nl
N6503 & 6 & 0.134 & 7.23 & 6.11 & 13.2 & 8.82 & 3 \nl
N6810 & 25 & 0.227 & 8.61 & 7.65 & 9.0  & 10.31 & 2 \nl
N6946 & 6 & 0.083 & 8.16 & 7.12 & 11.1 & 9.55 & 3 \nl
N7130 & 65\tablenotemark{c} & 0.473 & 9.21 & 8.31 & 8.0  & 11.05 & 5 \nl
N7552 & 20 & 0.212 & 8.72 & 7.56 & 14.2 & 10.63 & 2 \nl
N7582 & 18 & 0.143 & 8.50 & 7.28 & 16.7 & 10.39 & 2 \nl
IC342 & 4 & 0.044 & 7.81 & 6.76 & 11.1 & 9.32 & 3 \nl
IC2554 & 17 & 1.136 & 8.09 & 6.97 & 13.0 & 9.95\tablenotemark{f} & 2 \nl
U2855 & 20 & 0.126 & 8.52 & 7.44 & 11.9 & 10.40 & 2 \nl
Circinus & 4\tablenotemark{g} & 0.104 & 7.85 & 6.83 & 10.5 & 9.78\tablenotemark{h} & 5 \nl
Maffei2 & 3 & 0.950 & 7.64 & 6.68 & 9.0  & 9.33\tablenotemark{h} & 3 \nl
Arp220 & 74\tablenotemark{c} & 0.363 & 9.58 & $<$8.34 & $>$17.3 & 11.91 & 5 \nl
\nodata & \nodata & 0.242 & 9.72 & $<$8.39 & $>$21.8 & \nodata & 6 \nl
Arp299 & 42\tablenotemark{c} & 0.468 & 9.09 & $<$7.80 & $>$19.8 & 11.41 & 5 \nl
\nodata & \nodata & 0.312 & 8.99 & 7.68 & 20.7 & \nodata & 6 \nl
IRAS18293\tablenotemark{i} & 73\tablenotemark{j} & 2.389\tablenotemark{k} & 
9.79 & 8.56 & 16.8 & 11.46 & 2 \nl
\nodata & \nodata & 2.500\tablenotemark{k} & 9.86 & 8.44 & 25.9 & \nodata & 7 \nl
IRAS22132\tablenotemark{i} & 46\tablenotemark{c} & 0.320 & 9.07 & 8.12 & 8.9  & 10.84 & 7 \nl
ESO541-IG23  & 81\tablenotemark{c} & 1.604 & 9.59 & $<$8.36 & $>$17.0 & 11.36 & 2 \nl
\enddata
\tablenotetext{a}{1: Becker \& Freudling (1991), 2: Aalto et al. (1995),
3: Sage \& Isbell (1991), 4: Wiklind \& Henkel (1990),
5: Aalto et al. (1991), 6: Casoli et al. (1992b), 7: Garay et al. (1993),
8: Henkel et al. (1994)}
\tablenotetext{b}{Rice et al. (1988)}
\tablenotetext{c}{Estimated using a Hubble constant
$H_0$ = 75 km s$^{-1}$ Mpc$^{-1}$ with $V_{\rm GSR}$
given in de Vaucouleurs et al. (1991).}
\tablenotetext{d}{Sum of $D_{25}$ of both N4038 and N4039 is used.}
\tablenotetext{e}{Data in all regions (nuclus and four disk regions) are taken togegher
 (see Sage \& Isbell 1991).}
\tablenotetext{f}{Sanders et al. (1995)}
\tablenotetext{g}{Freeman et al. (1977)}
\tablenotetext{h}{Lonsdale et al. (1989)}
\tablenotetext{i}{IRAS18293 = IRAS 18293-3413 and IRAS22132 = IRAS 22132-3705}
\tablenotetext{j}{Strauss et al. (1992)}
\tablenotetext{k}{Optical diameter is taken from NED.}
\end{deluxetable}

\newpage

\begin{figure}
\epsfysize=15cm \epsfbox{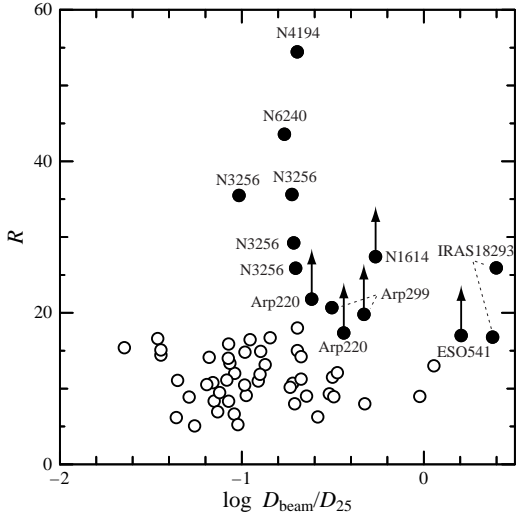}
\caption[]{
Diagram of $D_{\rm beam}/D_{25}$ vs. $R$.
Objects with $R \gtrsim 20$ (see the main text) are shown by filled circles
and labeled.
Note that IRAS18293 = IRAS 18293$-$3413 and ESO541 = ESO 541--IG 23.
\label{fig1}
}
\end{figure}

\begin{figure}
\epsfysize=15cm \epsfbox{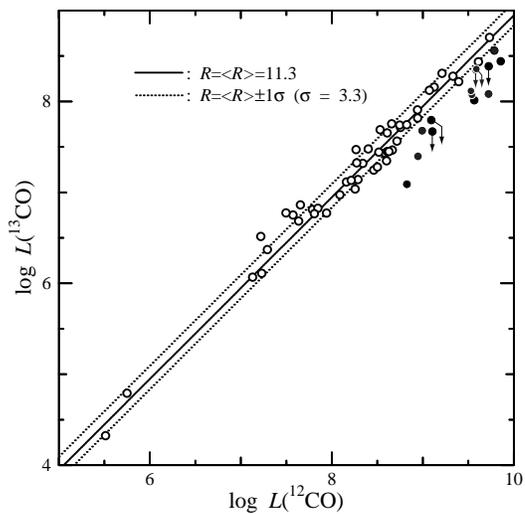}
\caption[]{
Diagram of $L$($^{12}$CO) vs. $L$($^{13}$CO).
The symbols have the same meaning as those in figure 1.
A solid line corresponds to the correlation for $R=<R>$ where $<R>$ is
the mean $R$ using the data of galaxies with $R<20$ ($<R>=11.3$).
The dotted lines corresponds to the correlation for $R=<R>\pm \sigma (R)$
where $\sigma (R) = 3.3$.
\label{fig2}
}
\end{figure}

\begin{figure}
\epsfysize=18.5cm \epsfbox{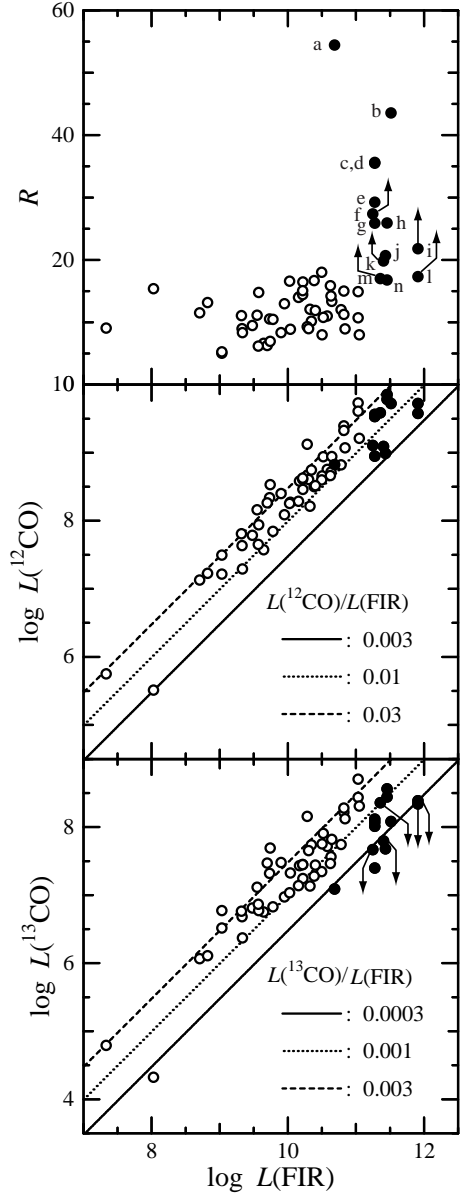}
\caption[]{
Diagrams of $R$ (top), $L$($^{12}$CO) (middle),
and $L$($^{13}$CO) (bottom) against $L$(FIR).
The symbols have the same meaning as those in Figure 1.
Alphabets in the top panel mean;
a: NGC 4194, b: NGC 6240, c: NGC 3256, d: NGC 3256, e: NGC 3256, f: NGC 1614,
g: NGC 3256, h: IRAS 18293$-$3413, i: Arp 220, j: Arp 299, k: Arp 299,
l: Arp 220, m: ESO 541--IG 23, and  n: IRAS 18293$-$3413.
\label{fig3}
}
\end{figure}

\end{document}